\documentclass[11pt,a4paper]{article}
\usepackage{authblk}
\usepackage{nameref}
\usepackage{hyperref}
\usepackage{graphicx}
\usepackage{subcaption}
\usepackage{mathrsfs}
\usepackage{mathtools}
\usepackage{amssymb}
\usepackage{here}

\DeclareMathOperator*{\argmax}{arg\,max}

\newcommand\Tstrut{\rule{0pt}{2.6ex}}         
\newcommand\Bstrut{\rule[-0.9ex]{0pt}{0pt}}   

\title{Machine learning for structure-guided materials and process design}

\author[1]{Lukas Morand*}
\author[2,3]{Tarek Iraki}
\author[3,4]{Johannes Dornheim}
\author[2]{Stefan Sandfeld}
\author[3]{Norbert Link}
\author[1]{Dirk Helm}

\affil[1]{Fraunhofer Institute for Mechanics of Materials IWM, Freiburg, Germany}
\affil[2]{Institute for Advanced Simulations - Materials Data Science and Informatics (IAS-9), Forschungszentrum J\"ulich GmbH, 52425, J\"ulich, Germany}
\affil[3]{Intelligent Systems Research Group ISRG, University of Applied Sciences, Karlsruhe, Germany}
\affil[4]{Institute for Applied Materials - Computational Materials Sciences IAM-CMS, Karlsruhe Institute of Technology, Karlsruhe, Germany} 
\affil[*]{corresponding author, email: lukas.morand@iwm.fraunhofer.de}

\begin{document}

\maketitle

\begin{abstract}
In recent years, there has been a growing interest in accelerated materials innovation in the context of the process-structure-property chain. In this regard, it is essential to take into account manufacturing processes and tailor materials design approaches to support downstream process design approaches. As a major step into this direction, we present a holistic and generic optimization approach that covers the entire process-structure-property chain in materials engineering. Our approach specifically employs machine learning to address two critical identification problems: a materials design problem, which involves identifying near-optimal material microstructures that exhibit desired properties, and a process design problem that is to find an optimal processing path to manufacture these microstructures. Both identification problems are typically ill-posed, which presents a significant challenge for solution approaches. However, the non-unique nature of these problems offers an important advantage for processing: By having several target microstructures that perform similarly well, processes can be efficiently guided towards manufacturing the best reachable microstructure. The functionality of the approach is demonstrated at manufacturing crystallographic textures with desired properties in a simulated metal forming process.

\end{abstract}

\section{Introduction}

\subsection{Motivation}

Accelerated materials innovation has become a core research field in integrated computational materials engineering (ICME) and is pushed forward strongly in materials science and engineering (cf. materials genome initiative \cite{de2019new}, European advanced materials initiative \cite{EU2030roadmap}). Essentially, the properties of such new sustainable, resilient, and high-performance materials depend on the microstructure of the material, which, in turn, depends on the manufacturing process. Consequently, designing new materials without taking into account the manufacturing process does not add significant value \cite{Grant2013new}. For an application in industry, groundbreaking technologies for modeling and optimization that cover the entire process-structure-property chain are required. 

\begin{figure}
\centering
\includegraphics[width=0.6\textwidth]{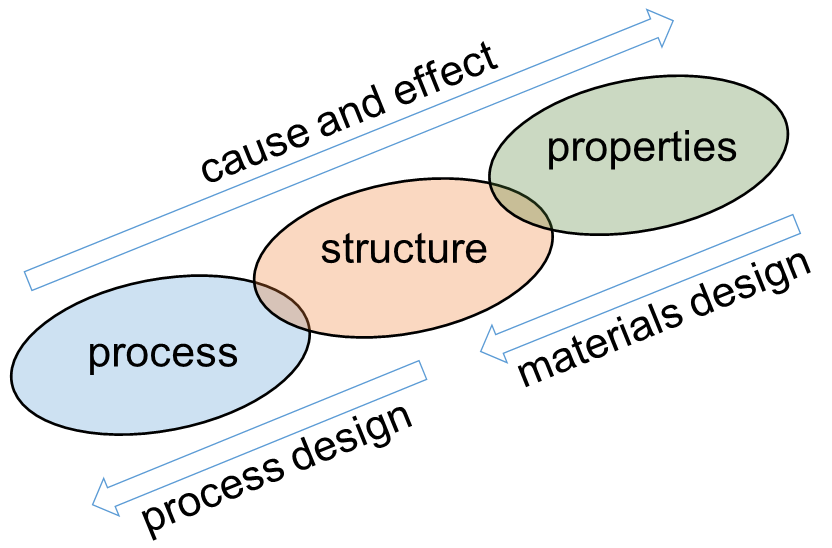}
\caption{Process-structure-property chain following \cite{olson1997computational}}
\centering
\label{fig:psp_chain}
\end{figure}

The process-structure-property chain, as depicted in Figure \ref{fig:psp_chain}, was originally introduced in \cite{olson1997computational} and describes fundamental relations in materials processing. A basic characteristic of this chain is its modularity: Each individual link represents a specific identification problem, which includes identifying microstructures for given desired properties (materials design) and finding optimal processing paths for targeted microstructures (process design). These identification problems are typically not well-posed in the sense of Hadamard \cite{hadamard1902problemes}, which presents a significant challenge for solution approaches \cite{agrawal2016perspective}. However, the non-unique nature of these problems offers an important advantage for processing: It enables a more flexible production as processes can be efficiently guided to manufacture the best reachable microstructure from a set of equivalent microstructures with respect to their properties.

Leveraging this advantage presents certain challenges that require specialized yet generic materials and process design approaches. These approaches must be capable of identifying multiple solutions to their individual design problems and must work in conjunction with each other. In an effort to address this challenge, the present paper proposes a solution by combining two recently developed machine learning approaches, as described in \mbox{\cite{dornheim2020structureguided}} and \mbox{\cite{Iraki.2021}}. The motivation behind using these approaches is elaborated in the following.

To handle the complexity and dimensionality of the identification problems, the use of machine learning has shown to be suitable for materials and process design applications \cite{liu2017materials, mozaffar2022mechanistic}.
In this work, a novel machine learning framework is introduced that combines the reinforcement learning-based process design approach developed in \mbox{\cite{dornheim2020structureguided}}: multi-equivalent goal structure-guided processing path optimization (MEG-SGGPO), with the materials design approach developed in \mbox{\cite{Iraki.2021}}: Siamese multi-task learning-based optimization (SMTLO). The framework is tailored to the specifics of process-structure-property optimization problems and, therefore, constitutes a significant advancement towards accelerated process and materials design. 

Although each of the approaches has been shown to work well for their individual design problems, the combined usage of both is not investigated yet. This is particularly the aim of the present work, while the approaches are enhanced by using a recently developed novel distance measure for one-point statistics microstructure representations: the Sinkhorn distance $\mathscr{D}_\mathrm{sh}$ \cite{iraki2022sinkhorn}. Specifically, the Sinkhorn distance highly suitable as it takes into account neighborhood information encoded in histogram-based microstructure representations. The importance of the distance measure can be seen in Figure \ref{fig:general_concept}, which depicts the general concept of the approach. In this work, we demonstrate the approach at manufacturing metallic materials with desired elastic and anisotropy properties, which are affected by the crystallographic texture that evolves during forming.

\subsection{Related work}

Pioneering work has been done in the field of materials and process design, however, typically either focusing on solving materials design problems (without taking into account processing) or focusing on solving process design problems for given desired properties (not taking into account the microstructure of materials) \mbox{\cite{kalidindi2004microstructure, sundararaghavan2005synergy, li2007processing, fullwood2010microstructure, adams2012microstructure}}. In the following, we discuss works that describe approaches for solving materials and process design problems with an emphasize on the application of machine learning.

A widely-used approach for designing materials is the microstructure sensitive design (MSD) approach  \cite{fullwood2010microstructure}. This approach primarily focuses on identifying microstructures that exhibit desired properties, with six out of the seven MSD steps being dedicated to this task. The processing of microstructures is only briefly addressed by the final MSD step. To the authors knowledge, there are only few works that show how to set up process-structure-property linkages for crystallographic texture optimization within the framework of MSD. One approach involves modeling crystallographic texture evolution as fluid flow in the orientation space. On this basis, so-called processing streamlines are used to guide from one point to another \cite{fullwood2010microstructure,li2007processing}. 

For example, in the case of optimizing the crystallographic texture of an orthotropic plate, these streamlines were calculated for the processing operations of tension and compression, and used to guide from a random crystallographic texture to required crystallographic textures \cite{adams2012microstructure}. These required crystallographic textures have been identified in beforehand based on the MSD approach \cite{kalidindi2004microstructure}. Alternatively, so-called texture evolution networks have been developed within the context of MSD, using a priory sampled processing paths to create a directed tree graph where microstructures are represented by the nodes on the graph \cite{Shaffer2010}. Graph search algorithms are then used to find optimal processing paths from an initial to a targeted microstructures.

Besides the MSD approach, other approaches exist that solve crystallographic texture optimization problems in terms of materials design, however, without solving a corresponding optimal processing problem. For instance, Kuroda and Ikawa \cite{kuroda2004texture} used a genetic algorithm to identify optimal combinations of typical fcc rolling (crystallographic) texture components for given desired properties. Also, Liu et al. \cite{liu2015predictive} used optimization algorithms, but incorporated machine learning techniques to efficiently identify significant features and regions of the orientation space. Surrogate-based optimization has also been explored in several studies, such as for handling uncertainties in materials design \cite{balachandran2016adaptive}. Alternatively, probabilistic modeling approaches can be used to directly to solve the inverse identification problem \cite{tran2021solving}.

Regarding process design, machine learning-based and data mining-based approaches have been proposed by several works, however, without solving the corresponding materials design problem in beforehand. One approach involves the use of principle component analysis to represent one-step and two- to three-step deformation processes and to identify the deformation sequences required for reaching target crystallographic textures \cite{acar2016linear,acar2018reduced}. Alternatively, a database approach can be used that stores microstructure representations and corresponding processing paths \cite{sundararaghavan2005synergy}. The database can be searched for desired crystallographic textures yielding optimal process paths. These process paths can then be fine-tuned using gradient-based optimization.
Another option is to store microstructure representations in a lower dimensional feature space generated by a variational autoencoder \cite{sundar2020database}.
In this lower dimensional feature space, optimal processing paths can be identified using a suitable distance measure. A more recent work for process design in terms of crystallographic texture, although optimizing process-property linkages, is described in \mbox{\cite{lin2023neural}}. Therein, neural networks are used to learn texture evolution in a deformation process, while an optimizer is used to drive the process to produce a material with desired properties. In this framework, however, crystallographic texture is varied only indirectly and it cannot be guaranteed that an optimal texture is found in terms of material properties and reachability by the process.

Recent research directions in the field of materials and process design are the usage of probabilistic methods such as Bayesian approaches to tackle the ill-posedness of the inverse problem. In \mbox{\cite{generale2024inverse}} a Bayesian methodology was used to estimate a posterior distribution of microstructures that are conditioned by a user-defined target property given a prior distribution. A similar approach was used in \mbox{\cite{generale2023bayesian}}, in which optimal process parameters were estimated for manufacturing a material with given desired material properties (process-property linkage). In terms of process-structure design problems, in \mbox{\cite{tran2020active}}, a Bayesian active learning approach is proposed that is based on Gaussian processes. While Bayesian methods generally suit well for solving such inverse problems, it can be hard to train these when the amount of data increases. Therefore, active learning approaches were developed that do not use Gaussian processes but instead make use of a committee-based active learning approach \mbox{\cite{morand2022efficient}}. Other novel approaches in materials engineering make use of knowledge graphs for microstructure representation and the application of graph neural networks for property prediction, as demonstrated in \mbox{\cite{thomas2023materials}}. Such approaches are rather descriptor-based approaches, such as the ones presented in \mbox{\cite{kumar2020inverse}} and \mbox{\cite{rassloff2024inverse}} for meta materials design.

Other recent approaches for designing materials make use of generative models, such as in \mbox{\cite{sandfeld2023generative}}. Therein a generative adversarial network is modified to generate microstructures based on continuous variables, which is an important prerequisite for designing materials. In \mbox{\cite{heyrani2021pcdgan}}, a further developed continuous conditioned GAN for diverse outputs was introduced and applied to a synthetic and real-world airfoil design task. Alternatively, in \mbox{\cite{wijaya2024analyzing}} a diffusion-based machine learning model is used for microstructure reconstruction with respect to given material properties. An alternative, descriptor-based approach for microstructure reconstruction can be found in \mbox{\cite{seibert2022descriptor}}. Furthermore, since the work by \mbox{\cite{sundar2020database}}, variational auto-encoder-based approaches gain popularity, such as in \mbox{\cite{fallani2024inverse}}  for the design of quantum-mechanical properties in small organic molecules. Recent developments in the design of molecules investigate the potential of using large language models, such as in \mbox{\cite{jablonka2024leveraging}}.

In the future, it is envisioned that design approaches will be implemented as applications on data platforms, as described in \mbox{\cite{sivan2024advances}}. Such platforms, like the one described in \mbox{\cite{nahshon2024semantic}}, enable the storage of heterogeneous material data in a FAIR manner, ensuring that the data is findable, accessible, interoperable, and reusable \mbox{\cite{scheffler2022fair}}. This, in turn, allows for (automated) data processing using applications, such as based on simulation or machine learning. The utilization of these platforms opens up the possibility of employing design approaches for experimentation, which has the potential to lead to autonomous materials discovery and manufacturing \mbox{\cite{bukkapatnam2023autonomous}}.

\subsection{Structure of this work}
In the following Section \ref{sec:methods}, we introduce the structure-guided materials and process design approach as well as the metal forming process simulation at which the functionality of the approach is demonstrated and the used representation of crystallographic texture. Afterwards, in Section \ref{sec:results}, the results for optimizing the crystallographic texture in the metal forming process for given desired properties are shown. The results are discussed in Section \ref{sec:discussion} and an outlook is given in Section \ref{sec:conclusion}.

\section{Methods} 
\label{sec:methods}

The methods section is split into two parts: First, we present the basics for the structure-guided materials and process design approach studied in this paper, and, second, we present the domain specific application case at which we evaluate the presented approach.

\subsection{Structure-guided materials and process design}

\subsubsection{General concept}

\begin{figure}[H]
\includegraphics[width=\textwidth]{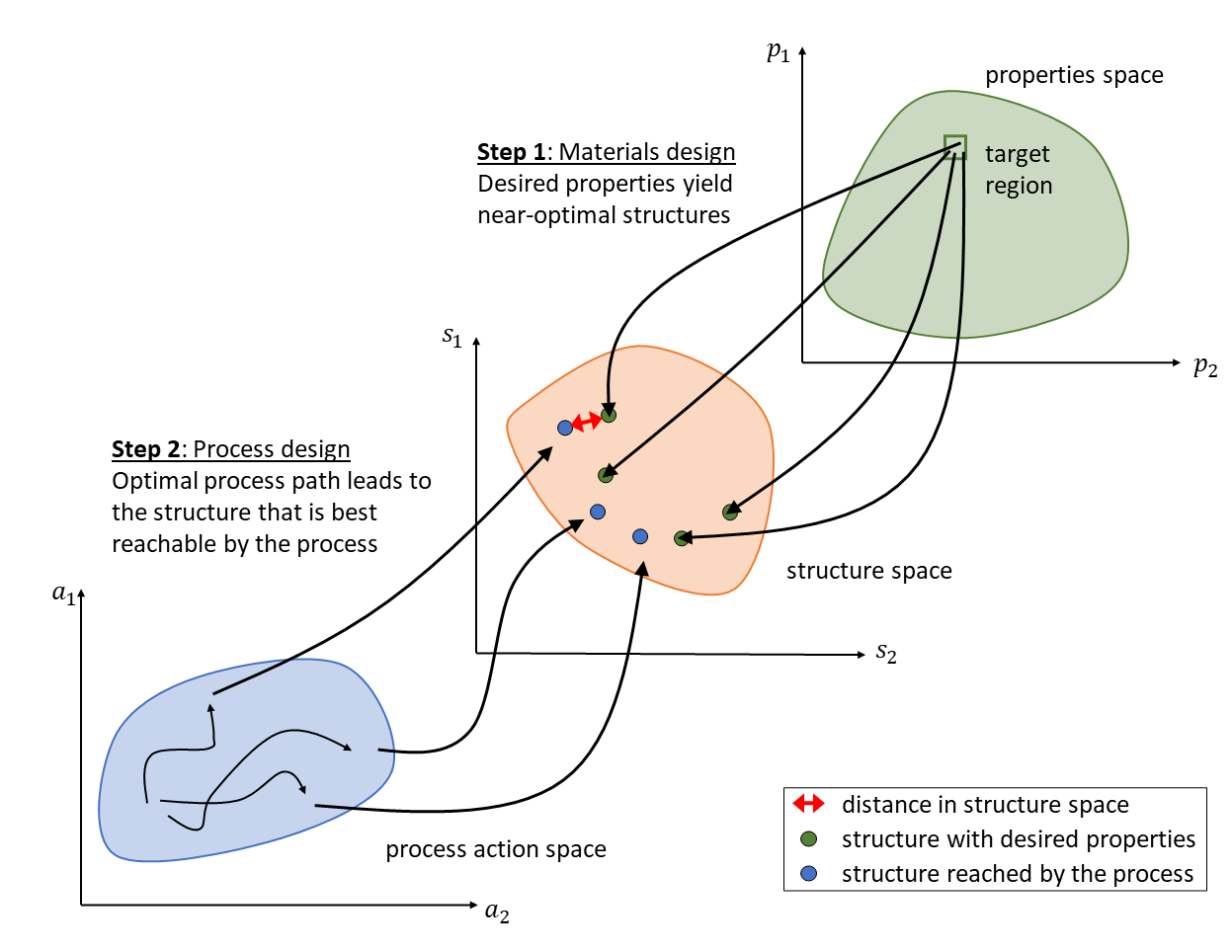}
\caption{General concept for structure-guided materials and process design. The first step (materials design) is addressed by the SMTLO approach, while the second step (process design) is addressed by the MEG-SGGPO approach.}
\centering
\label{fig:general_concept}
\end{figure}

The approach presented in this paper combines two machine learning approaches for materials design, namely Siamese multi-task learning-based optimization (SMTLO), and for process design, namely multi-equivalent goal structure guided processing path optimization (MEG-SGGPO), see Figure \ref{fig:general_concept}. The approach starts in the properties space, where a target region of desired properties is defined. Then, the SMTLO approch is applied to identify a set of diverse microstructures that yield material properties inside the target region (Step 1). When having identified this set, MEG-SGGPO identifies the process path that leads to the best reachable microstrucutre (Step 2). The microstructure space is linking process and properties, and, therefore, is of overall importance also for our approach. A suitable distance measure is one of the core ingredients for the approach to work. The proposed approach is generic in nature, making it applicable to any kind of process-structure-property relations. Furthermore, it is not limited to particular input/output data formats, and can even be applied to image data.

\subsubsection{Siamese multi-task learning-based optimization (SMTLO)}
\label{sec:methods_SMTL}

\begin{figure}
\includegraphics[width=\textwidth]{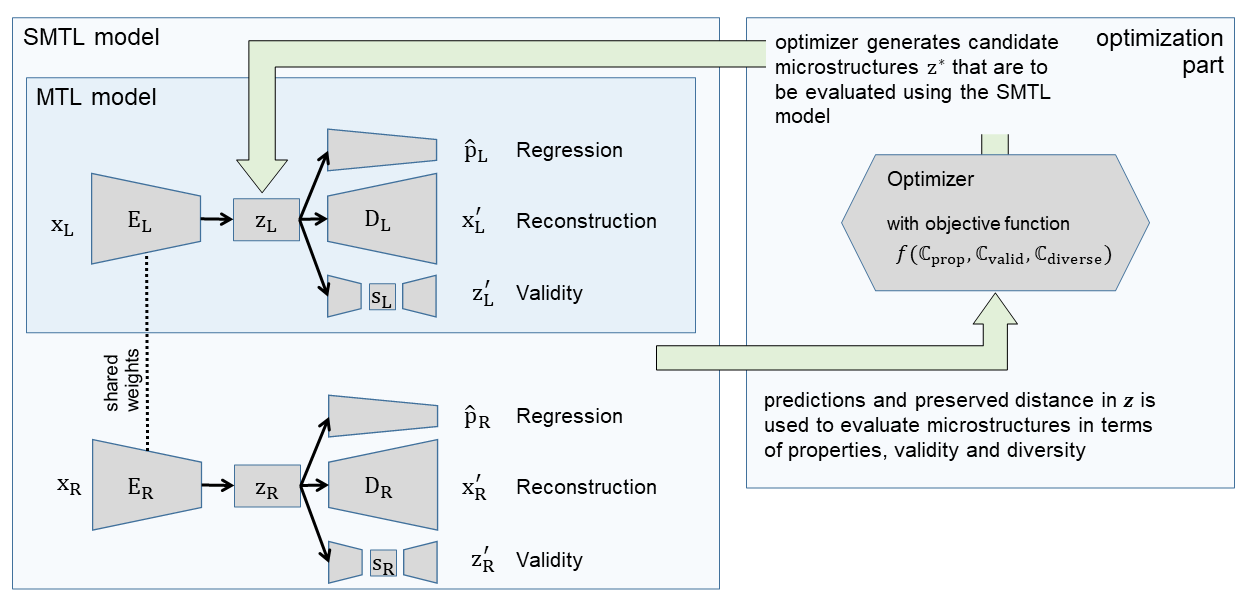}
\caption{The neural networks-based Siamese multi-task learning model (left) and the optimization part (right). Together, both parts build the SMTLO approach to solve materials design probplems.}
\centering
\label{fig:SMTLO}
\end{figure}

The SMTLO approach described in \cite{Iraki.2021} consists of a Siamese neural networks-based multi-task learning model and an optimizer. The latter uses the predictions of the learned neural network model to generate candidate microstructures that are supposed to (i) yield properties inside a defined target region and (ii) are reachable by the underlying manufacturing process, see Figure \ref{fig:SMTLO}. The details of the SMTLO approach is described in the following. Alternative approaches, that focus only the mapping of microstructures to properties can be found, for example in \cite{nguyen2024efficient}.

The multi-task learning model is grounded on an encoder $f_\mathrm{enc}$ that transforms a microstructure representation $\boldsymbol{x}$ into a lower dimensional latent feature space representation $\boldsymbol{z}$ 
\begin{equation}
\boldsymbol{z} = f_\mathrm{enc}(\boldsymbol{x},\boldsymbol{\theta}_\mathrm{enc}),
\end{equation}
with the trainable parameters $\boldsymbol{\theta}_\mathrm{enc}$.
The encoder is equipped with three heads, each of which is designed to solve a specific learning task. These are
\begin{enumerate}
\item a decoder head that reconstructs the microstructure representations $\boldsymbol{x}$ from $\boldsymbol{z}$
\begin{equation}
\boldsymbol{x}^\prime = f_\mathrm{dec}(\boldsymbol{z},\boldsymbol{\theta}_\mathrm{dec}),
\end{equation}
with the trainable parameters $\boldsymbol{\theta}_\mathrm{dec}$. The encoder-decoder part is trained using a loss term that minimizes the Sinkhorn distance between the original microstructure $\boldsymbol{x}$ and its reconstruction $\boldsymbol{x}^\prime$ \cite{iraki2022sinkhorn}
\begin{equation}
\mathscr{L}_{\mathrm{recon}} = \mathscr{D}_{\mathrm{sh}} (\boldsymbol{x},\boldsymbol{x}^\prime).
\end{equation}

\item a prediction head $f_\mathrm{regr}$ to infer material properties $\hat{\boldsymbol{p}}$
\begin{equation}
\hat{\boldsymbol{p}} = f_\mathrm{regr}(\boldsymbol{z},\boldsymbol{\theta}_\mathrm{regr}),
\end{equation}
with the trainable parameters $\boldsymbol{\theta}_\mathrm{regr}$. The sample-wise loss term is defined by the mean squared error between predicted properties $\hat{\boldsymbol{p}}$ and true properties $\boldsymbol{p}$:
\begin{equation}
\mathscr{L}_{\mathrm{regr}} = \frac{1}{n_\mathrm{p}}\sum_i^{n_\mathrm{p}} (p_i - \hat{p}_i)^2,
\end{equation}
with the number of properties $n_\mathrm{p}$.

\item an auto-encoder head that serves as an anomaly detector to estimate whether a microstructure in its latent representation belongs to the set of known microstructures (defined by the training data) or not:
\begin{equation}
\boldsymbol{z}^\prime = f_\mathrm{valid}(\boldsymbol{z},\boldsymbol{\theta}_\mathrm{valid}),
\label{eq:validity}
\end{equation}
with the trainable parameters $\boldsymbol{\theta}_\mathrm{valid}$. 
For this auto-encoder, a mean squared error loss function is used:
\begin{equation}
\mathscr{L}_{\mathrm{valid}} 
= \frac{1}{n_\mathrm{z}}\sum_i^{n_\mathrm{z}} (z_i - z^\prime_i)^2,
\end{equation}
with $n_\mathrm{z}$ dimensions of the latent feature space.

\end{enumerate}

The neural networks-based multi-task learning model that is used to solve the above mentioned learning tasks is trained using a combined loss function
\begin{equation}
\mathscr{L}_\mathrm{MTL} = \mathscr{W}_{\mathrm{recon}} \mathscr{L}_{\mathrm{recon}} + \mathscr{W}_{\mathrm{regr}} \mathscr{L}_{\mathrm{regr}} + \mathscr{W}_{\mathrm{valid}} \mathscr{L}_{\mathrm{valid}},
\end{equation}
with individually weighted loss terms using the parameters $\mathscr{W}_{\mathrm{recon}}$, $\mathscr{W}_{\mathrm{regr}}$ and $\mathscr{W}_{\mathrm{valid}}$. 

The objective of the materials design step is to enable the optimizer to identify a diverse set of microstructures for given desired material properties, each microstructure being reachable by the underlying manufacturing process. In order to quantify diversity, a distance measure is required in the latent feature space. To ensure that the original microstructure distance is preserved in the lower dimensional latent feature space, the multi-task learning model is embedded in a Siamese neural network model \cite{bromley1993signature}. This involves training two twin models simultaneously, with shared weights in the encoder, and adding a further loss term that applies to the latent feature space. Note that the two twin models are trained using different input and output vectors, denoted with the subscripts $_L$ and $_R$ in the following. The preservation loss leads to multidimensional scaling (see \cite{Kruskal1964MultidimensionalSB} and \cite{Multidimensional-Scaling-Cox2008}) and writes
\begin{equation}
\mathscr{L}_{\mathrm{pres}} = (\mathscr{D}_{\mathrm{sh}}(\boldsymbol{x}_{L},\boldsymbol{x}_{R})-\mathrm{dist}(\boldsymbol{z}_{L},\boldsymbol{z}_{R}))^2,
\end{equation}

with the absolute distance in the latent feature space 
\begin{equation}
\label{eq:distance-latent-space}
\mathrm{dist}(\boldsymbol{z}_{L},\boldsymbol{z}_{R}) = \frac{1}{n_\mathrm{z}} \sum_i^{n_\mathrm{z}} |z_{L,i} - z_{R,i}|.
\end{equation}
The subscripts $_L$ and $_R$ are used to distinguish between the two simultaneously trained twins parts. The overall loss function writes
\begin{equation}
\label{eq:smtl-loss}
\mathscr{L} = \mathscr{W}_{\mathrm{recon}} \mathscr{L}_{\mathrm{recon}} + \mathscr{W}_{\mathrm{regr}} \mathscr{L}_{\mathrm{regr}} + \mathscr{W}_{\mathrm{valid}} \mathscr{L}_{\mathrm{valid}} + \mathscr{W}_{\mathrm{pres}} \mathscr{L}_{\mathrm{pres}} + \lambda \Omega(\boldsymbol{\theta}),
\end{equation}
with the regularization term $ \Omega(\boldsymbol{\theta})$, and the weight $\mathscr{W}_{\mathrm{pres}}$ for the preservation loss.

After training the Siamese multi-task learning model, we use a genetic optimizer \cite{zhang2009JADE} to search for candidate microstructures in the obtained latent feature space. The optimizer minimizes three cost terms:
$\mathscr{C}_{\mathrm{prop}}, \mathscr{C}_{\mathrm{valid}}, \mathscr{C}_{\mathrm{divers}}$, which are explaing in detail in \cite{Iraki.2021}. Briefly described, $\mathscr{C}_{\mathrm{prop}}$ describes the distance between the predicted properties $\hat{\boldsymbol{p}}$ for a candidate microstructure $\boldsymbol{z}^*$ to the target region. $\mathscr{C}_{\mathrm{valid}}$ aims to penalize candidate microstructures that are not inside the region of known microstructures. $\mathscr{C}_{\mathrm{divers}}$ aims to enforce candidate to microstructures be as far away as possible from each other in the actual population (in other words as diverse as possible). 
The overall objective function writes
\begin{equation}
\label{eq:matdesign_obj}
\mathscr{C} =  \mathscr{V}_{\mathrm{prop}} \mathscr{C}_{\mathrm{prop}} + \mathscr{V}_{\mathrm{valid}} \mathscr{C}_{\mathrm{valid}} + \mathscr{V}_{\mathrm{divers}} (1 + \mathscr{C}_{\mathrm{divers}}),
\end{equation}
with individual weights $\mathscr{V}_{\mathrm{prop}}$, $\mathscr{V}_{\mathrm{valid}}$, and $\mathscr{V}_{\mathrm{divers}}$. 
The property cost term directs the candidate microstructures to align with the desired target properties. The validity cost ensures the optimizer stays within the region of valid microstructures, while the diversity cost guarantees sufficient variation among the candidate microstructures. By optimizing the objective function in Equation (\mbox{\ref{eq:matdesign_obj}}), only those candidate microstructures that meet the necessary criteria are selected for the process design step.

\subsubsection{Multi-equivalent goal structure-guided processing path optimization (MEG-SGGPO)}
\label{ss:meg-sggpo}

In reinforcement learning, an agent learns to take optimal decisions by interacting with its environment in a sequence of discrete time steps $t$. The MEG-SGGPO approach \cite{dornheim2020structureguided} is tailored for efficient learning in the case of multiple equivalent target microstructures. Given a set $\mathcal{G}$ of near-optimal microstructures $\check{g}^i \in \mathcal{G}$, an initial microstructure $g_0$, and a microstructure generating process $p(g_t, a_t)=g_{t+1}$, the go of the approach is to find a sequence of processing parameters $a_t\in A$ that guide the process towards manufacturing the best reachable microstructure $g^*\in\mathcal{G}$. 

MEG-SGGPO is based on deep Q-networks\cite{mnih2015human} and its extensions \cite{van2016deep,schaul2015prioritized,wang2015dueling}. Deep Q-networks are derived from Q-learning, where the so-called Q-function is approximated from data gathered by the reinforcement learning agent while following a policy  $a=\pi(g)$. The Q-function models the expected sum of future reward signals for state action pairs $(g_t, a_t)$. In MEG-SGGPO the Q-function is generalized to also take the goal microstructure into account and is, in its recursive form, defined as
\begin{equation}
	\label{eq_stateActionValueFunctionMG}
	\mathcal{Q}(g_t, a_t, \check{g}^i)=\mathbb{E}_{P, \pi}\Big[R(g_t, g_{t+1}, \check{g}^i)+ \max_{a_{t+1}\in A} \mathcal{Q}(g_{t+1}, a_{t+1}, \check{g}^i) \Big],
\end{equation}
where $R(g_t, g_{t+1}, \check{g}^i)$ is a per goal microstructure pseudo reward function
\begin{equation}
	\label{eq_reward}
	R(g_t, g_{t+1}, \check{g}^i)=\dfrac{1}{\mathscr{D}(g_{t+1}, \check{g}^i)}-\dfrac{1}{\mathscr{D}(g_t, \check{g}^i)},
\end{equation}
which is based on a microstructure distance function $\mathscr{D}$.

The process design task is a two-fold optimization problem to (i) identify the best reachable goal microstructure $\check{g}^{i*}$ and (ii) learn the optimal policy $\pi^*$. MEG-SGGPO addresses this by using the generalized functions defined above in a nested reinforcement learning procedure. At the beginning of each episode (i.e. an execution of the process during learning), the agent uses the generalized Q-function to identify the targeted microstructure $\check{g}^{i'}$ from $\mathcal{G}$ and determines the processing parameters during the episode. In a first step, an estimation of the best reachable goal microstructure $\check{g}^{i*}$ is extracted from the current $\mathcal{Q}$ estimation by
\begin{equation}
	\label{eq_goal_priorization3}
	\check{g}^{i*}=\argmax_{\check{g}^i \in\mathcal{G}}\Big[ \mathcal{V}(g_0, \check{g}^i)+\dfrac{1}{\mathscr{D}(g_{0}, \check{g}^i)}\Big],
\end{equation}
where the generalized state-value function $\mathcal{V}$ is defined as
\begin{equation}
	\label{eq_stateValueFunctionMG}
	\mathcal{V}(g, \check{g}^i)=\max_{a\in A} \mathcal{Q}(g, a, \check{g}^i).
\end{equation}
The targeted microstructure is chosen in an $\epsilon$-greedy approach \cite{sutton1992reinforcement}, where the agent chooses random targets and processing parameters in an $\epsilon$ fraction of the cases ($0 \leq \epsilon \leq 1$) and optimizes targets and processing parameters in the remaining cases. The optimal processing path is identified simultaneously in an inner loop by using the Q-learning approach. 

\subsection{Application case}

\subsubsection{Crystallographic texture optimization}

To study it's functionality, the structure-guided materials and process design approach is applied to a crystallographic texture optimization problem in a metal forming process. The process-structure-property chain adapts to cold forming operations that change the crystallographic texture of a material, which significantly affects elastic and plastic properties. 
The underlying metal forming process simulation is described in the following section, as well as the determination of the elastic and plastic properties that we are going to target.

We have chosen crystallographic texture optimization as the focal point of this paper for the following reasons:  Crystallographic texture of polycrystals and their corresponding properties is one of the core topics of materials engineering. In particular, understanding how microstructural features affect the elastic and anisotropic properties of metals plays an important role in manufacturing processes such as sheet metal forming. At the same time, crystallographic texture is a complex and high-dimensional microstructural feature, rendering the materials and process design problems inherently challenging. Therefore, it serves as an ideal benchmark problem to demonstrate the effectiveness of our proposed approach.

\subsubsection{Metal forming process simulation} \label{sec:meth_sim}

In this study, we use the metal forming process simulation as is described in  \cite{dornheim2020structureguided}. The simulation applies a deformation $\hat{\boldsymbol{F}}$
\begin{equation}
\hat{\boldsymbol{F}} = \boldsymbol{R} \widetilde{\boldsymbol{F}} \boldsymbol{R}^\top,
\end{equation}
with $\boldsymbol{R}$ being a rotation matrix that describes one out of 25 possible loading directions. The deformation $\widetilde{\boldsymbol{F}}$ is defined using orthogonal basis vectors $\boldsymbol{e}_i$ and the operator $\otimes$ for the dyadic product
\begin{equation}
\widetilde{\boldsymbol{F}} = \widetilde{F}_{11} \boldsymbol{e}_1 \otimes \boldsymbol{e}_1 + \widetilde{F}_{22} \boldsymbol{e}_2 \otimes \boldsymbol{e}_2 + \widetilde{F}_{33} \boldsymbol{e}_3 \otimes \boldsymbol{e}_3
\end{equation}
with $\widetilde{F}_{11}$ corresponding to $10\%$ strain increments. $\widetilde{F}_{22},\widetilde{F}_{33}$ are adjusted such that the stresses are in balance. The metal forming process consists of seven subsequent loading steps, each in a separate loading direction. For this study, we conducted 76980 random process paths to generate the training data for the SMTLO approach.

The underlying material model is a crystal plasticity model of Taylor-type \cite{Kalidindi.1992}. The volume averaged stress for $n_\mathrm{oris}$ crystals with different orientations is calculated by
\begin{equation}
\overline{\boldsymbol{T}} = \frac{1}{V} \sum_i^{n_\mathrm{oris}} \boldsymbol{T}^{(i)} V^{(i)},
\end{equation}
with the total volume $V$, the individual volume of each crystal $V^{(i)}$, and the Cauchy stress tensor $\boldsymbol{T}^{(i)}$.

With the multiplicative decomposition of the deformation gradient in its elastic and plastic part
\begin{equation}
\boldsymbol{F} = \boldsymbol{F}_\mathrm{e} \cdot  \boldsymbol{F}_\mathrm{p},
\end{equation}
and the conversion formula for the stress tensor in the intermediate configuration $\boldsymbol{T}^*$
\begin{equation}
\boldsymbol{T}^* = \boldsymbol{F}_\mathrm{e}^{-1} \cdot (\mathrm{det}(\boldsymbol{F}_\mathrm{e})\boldsymbol{T}) \cdot \boldsymbol{F}_\mathrm{e}^{-\top},
\end{equation}
the Cauchy stress tensor is derived using
\begin{equation}
\boldsymbol{T}^* = \frac{1}{2} \mathbb{C} : (\boldsymbol{F}_\mathrm{e}^\top \cdot \boldsymbol{F}_\mathrm{e} - \boldsymbol{I}),
\end{equation}
where $\boldsymbol{I}$ denotes the second order identity tensor and $\mathbb{C}$ the fourth order elastic stiffness tensor.

The evolution of the plastic deformation is described using the plastic part of the velocity gradient
\begin{equation}
\boldsymbol{L}_\mathrm{p} = \dot{\boldsymbol{F}_\mathrm{p}} \cdot \boldsymbol{F}_\mathrm{p}^{-1} = \sum_\alpha \dot{\gamma}^{(\alpha)} \boldsymbol{m}^{(\alpha)} \otimes \boldsymbol{n}^{(\alpha)},
\end{equation}
with the slip rates $\dot{\gamma}^{(\alpha)}$ on slip system $\alpha$ that is defined by the slip plane normal $\boldsymbol{n}^{(\alpha)}$ and the slip direction $\boldsymbol{m}^{(\alpha)}$. The slip rates are calculated by a phenomenological power law. The crystal reorientation is calculated by applying a rigid body rotation derived from the polar decomposition of $\boldsymbol{F}_\mathrm{e}$ to the original orientation, see \cite{ling2005numerical,iraki2022sinkhorn}. For the metal forming process simulation, the same material model parameters have been used as in \cite{dornheim2020structureguided}.

For the purpose of this study, we use the above described material model to evaluate the Young's modulus $E$ and an anisotropic property $\widetilde{R}$ in three orthogonal directions after a process run. The Young's modulus is calculated using the slope of the stress-strain curve in the elastic regime, that results after applying uniaxial tension. $\widetilde{R}$ is inspired by the Lankford coefficients in sheet metal forming and is calculated as the ratio between transverse strain and the strain in the loading direction in the plastic regime, also after applying uniaxial loading. 

\subsubsection{Crystallographic texture representation and texture distance measure} \label{sec:texture_repr}

Crystallographic texture is represented by using the histogram-based description introduced by Dornheim et al. \cite{dornheim2020structureguided}. The orientation histogram (here, we used a soft-assignment factor of 3) consists of 512 orientation bins that are nearly uniformly distributed in the cubic-triclinic fundamental zone. The bin orientations were created using the software \textit{neper} \cite{quey2011large, Quey2018}. 
In contrast to the original MEG-SGGPO and SMTLO approach,  in this work, crystallographic texture distances are measured using the Sinkhorn distance that is applied to the histogramm representations \cite{iraki2022sinkhorn}. Specifically, the Sinkhorn distance is an efficient implementation of the Earth Movers distance that measures the least amount of work necessary to transform one histogram into the other \cite{Rubner-2000-EMD-Image-Retrieval, Cuturi-2013-sinkhorn-distance}. Therefore, local orientation distances \cite{Huynh2009} encoded in the histograms are taken into account, in contrast to the originally proposed Chi-Squared distance, which is basically a bin-wise comparison of two histograms.

\section{Results}
\label{sec:results}

\subsection{Solving the materials and process design tasks}


The following demonstrates how the proposed approach can be implemented to solve a coupled materials and process design problem, specifically in relation to processing crystallographic textures with desired properties.
First, it is shown how near-optimal crystallographic textures are identified for given elastic and anisotropy properties using SMTLO. The properties are in particular the Young's moduli $E_i$ and anisotropy measures $\widetilde{R}_i$, both in three orthogonal directions. Once near-optimal crystallographic textures are identified, in a second step, MEG-SGGPO is used to guide the underlying manufacturing process to produce the best reachable crystallographic texture out of the set of identified near-optimal textures.

\begin{figure}
  \includegraphics[width=0.90\textwidth]{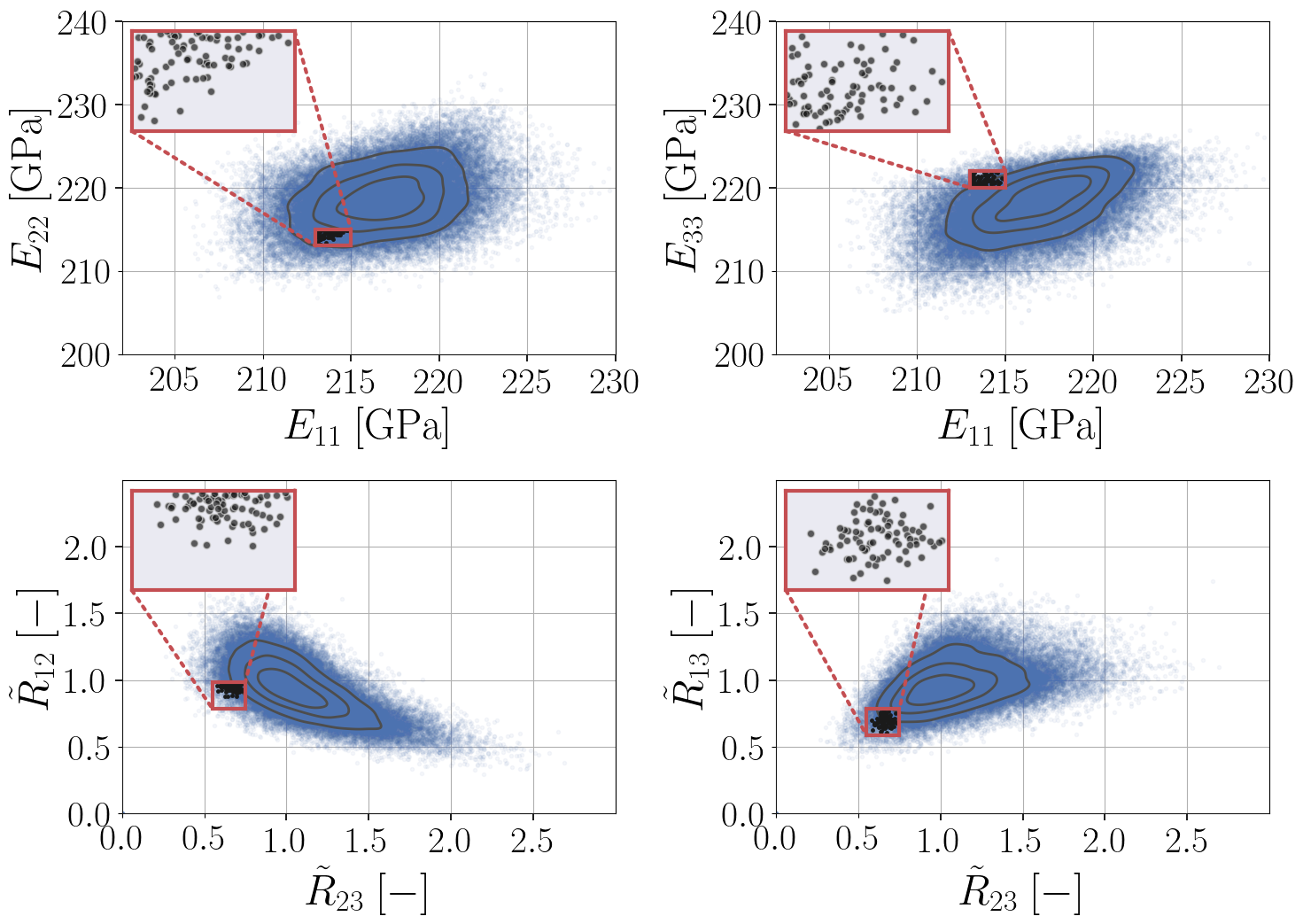}
\caption{Projections of the properties space showing the data distribution and the target region. The distribution of the underlying point cloud is displayed in blue, while the gray dots mark data points that are already located inside the target region. The black isolines indicate regions with the same point cloud density.}
\centering
\label{fig:res_matdesign_TR}
\end{figure}

The basis for applying the SMTLO approach is a data set of 76980 samples, composed of crystallographic textures and corresponding properties. 2D projections of the training data distribution, generated by the metal forming process simulation, are shown in Figure \ref{fig:res_matdesign_TR} including the region delineating the desired properties (target region). In this study, the target region is centered at $E_{11}, E_{22}, E_{33} = 214, 214, 221$ GPa and $\widetilde{R}_{23}, \widetilde{R}_{12}, \widetilde{R}_{13} = 0.65, 0.685, 0.885$. Its width equals $2$ GPa for $E_i$ and $0.2$ for $\widetilde{R}_i$. Both $E_i$ and $\widetilde{R}_i$ are determined in three orthogonal directions at the material point at the end of the process. $E_i$ is given by the slope in the elastic regime, while $\widetilde{R}_i$ is calculated by the ratio between transverse and tensile strain in the plastic regime when applying uniaxial tension. In Figure \ref{fig:res_matdesign_TR}, we highlighted training data points that are already located inside the target region. We use this set of $80$ crystallographic textures as benchmark set.

\subsection{Materials design using SMTLO} \label{sec:res_matdesign}
The Siamese multi-task learning model is realized via feedforward neural networks with $tanh$ activation functions and are implemented based on the Pytorch API \cite{Pytorch-2019}. For hyperparameter optimization the random search method \cite{Bergstra2012RandomSearch} is utilized using 5-fold cross-validation. The training data set comprises 80\% of the total data, while the test data set accounts for the remaining 20\%. The following hyperparameters have been chosen: The Glorot Normal method \cite{Glorot2010WeightInit} is used for weight initialization and the Adam optimizer \cite{Kingma2015Adam} is used with the following parameters: learning rate = 0.001, weight decay = $10^{-6}$, batch size = 128. 

\begin{figure}
  \includegraphics[width=0.90\textwidth]{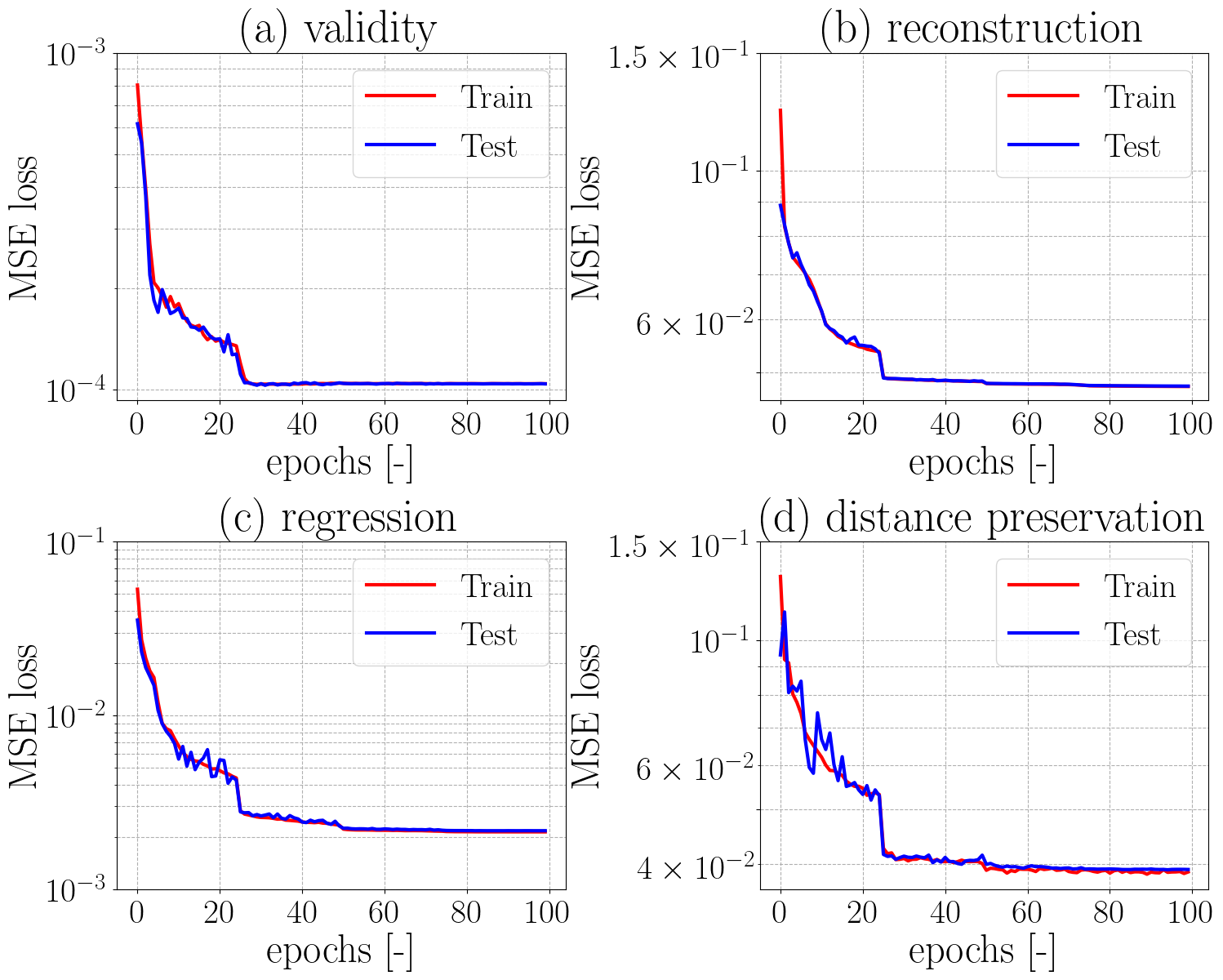}
\caption{Loss curves of the Siamese multi-task learning model for the test dataset (blue) and the training dataset (red) across tasks: (a) validity, (b) reconstruction, (c) regression, and (d) distance preservation, plotted on a logarithmic scale.}
\centering
\label{fig:smtl-loss-curves}
\end{figure}

The Siamese multi-task learning model is trained for 100 epochs, where the best intermediate result of the test set is retained to prevent over fitting and to apply early stopping \cite{Prechelt2012EarlyStopping}. Before the model is trained, the loss terms are scaled to values between 0 and 1 in order to make them comparable. The following weights for the scaled loss terms in Eq. (\ref{eq:smtl-loss}) were based on hyper parameter optimization: $\mathscr{W}_{\mathrm{recon}}=0.2$, $\mathscr{W}_{\mathrm{regr}}=0.06$, and $\mathscr{W}_{\mathrm{valid}}=0.04$, and $\mathscr{W} _{\mathrm{pres}}=0.70$. 
The loss curves resulting from the model training are collected for both, the test and the training dataset, for each individual epoch and are shown for each task separately in Figure \mbox{\ref{fig:smtl-loss-curves}}. It can be seen that for the validity, reconstruction, and regression tasks, the loss is minimized up to epoch 50 with no signs of overfitting. For the distance preservation task, the loss continues to decrease beyond epoch 50, also without any overfitting observed. For the cost function of the optimizer (defined in Eq. (\ref{eq:matdesign_obj})) that we need after having a trained Siamese multi-task learning model, we use the weights $\mathscr{V}_{\mathrm{prop}}=0.75$, $\mathscr{V}_{\mathrm{valid}}=0.05$, and $\mathscr{V}_{\mathrm{divers}}=0.2$. 

The results for the properties prediction are given by the the mean absolute error (MAE) between the true and predicted Young’s moduli and $\widetilde{R}$-values: $\mathrm{MAE}_\mathrm{E}$ = 0.368 [GPa] and $\mathrm{MAE}_\mathrm{\widetilde{R}}$ = 0.032 [-]. The result of the distance preservation is measured by the coefficient of determination $R^2$, between the Sinkhorn distance of two input crystallographic textures and the $l_1$ distance of their corresponding latent feature vectors: $R^2(\mathscr{D}_{\mathrm{sh}} (\boldsymbol{x}_L,\boldsymbol{x}_R), l_1(\boldsymbol{z}_{L},\boldsymbol{z}_{R})) = 90.22 [\%].$

\begin{figure}
\centering
\includegraphics[width=0.80\linewidth]{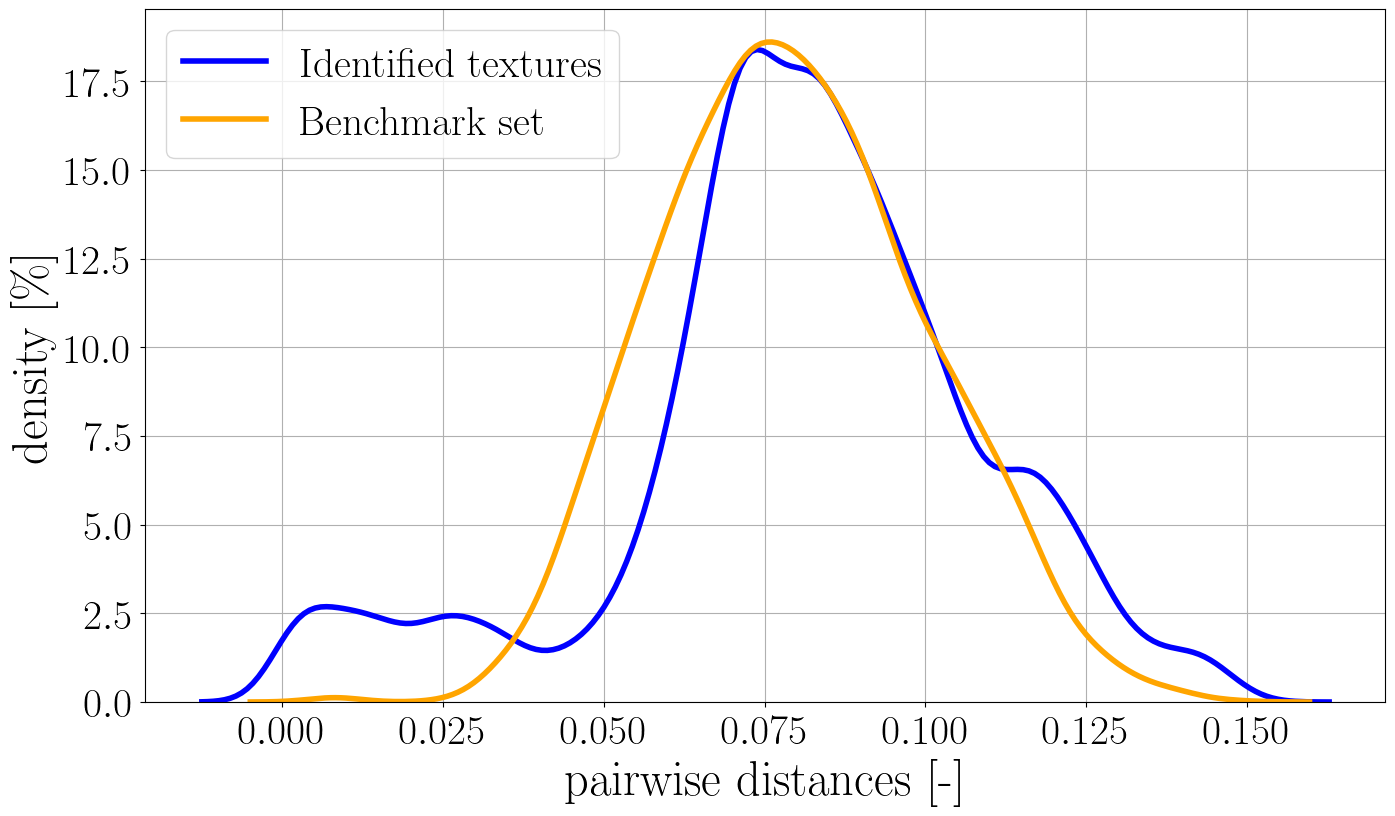}
\caption{Density of pairwise distances between crystallographic textures belonging to the set of identified crystallographic textures (red) and between crystallographic textures belonging to the benchmark set (green), evaluated in the latent space of the Siamese multi-task learning model}
\label{fig:res_matdesign_divplot}
\end{figure}

Using the SMTLO approach (with the trained multi-task learning model as described above), we were able to identify a set of 175 near-optimal crystallographic textures. As we aim to identify a preferably diverse set of crystallographic textures, we depict their pairwise distances in Figure \ref{fig:res_matdesign_divplot} and compare them to the benchmark set. As one can easily see, the SMTLO approach is able to find a set of crystallographic textures that differ more to each other than the benchmark set. 

\begin{figure}
	\centering
	\includegraphics[width=0.99\textwidth]{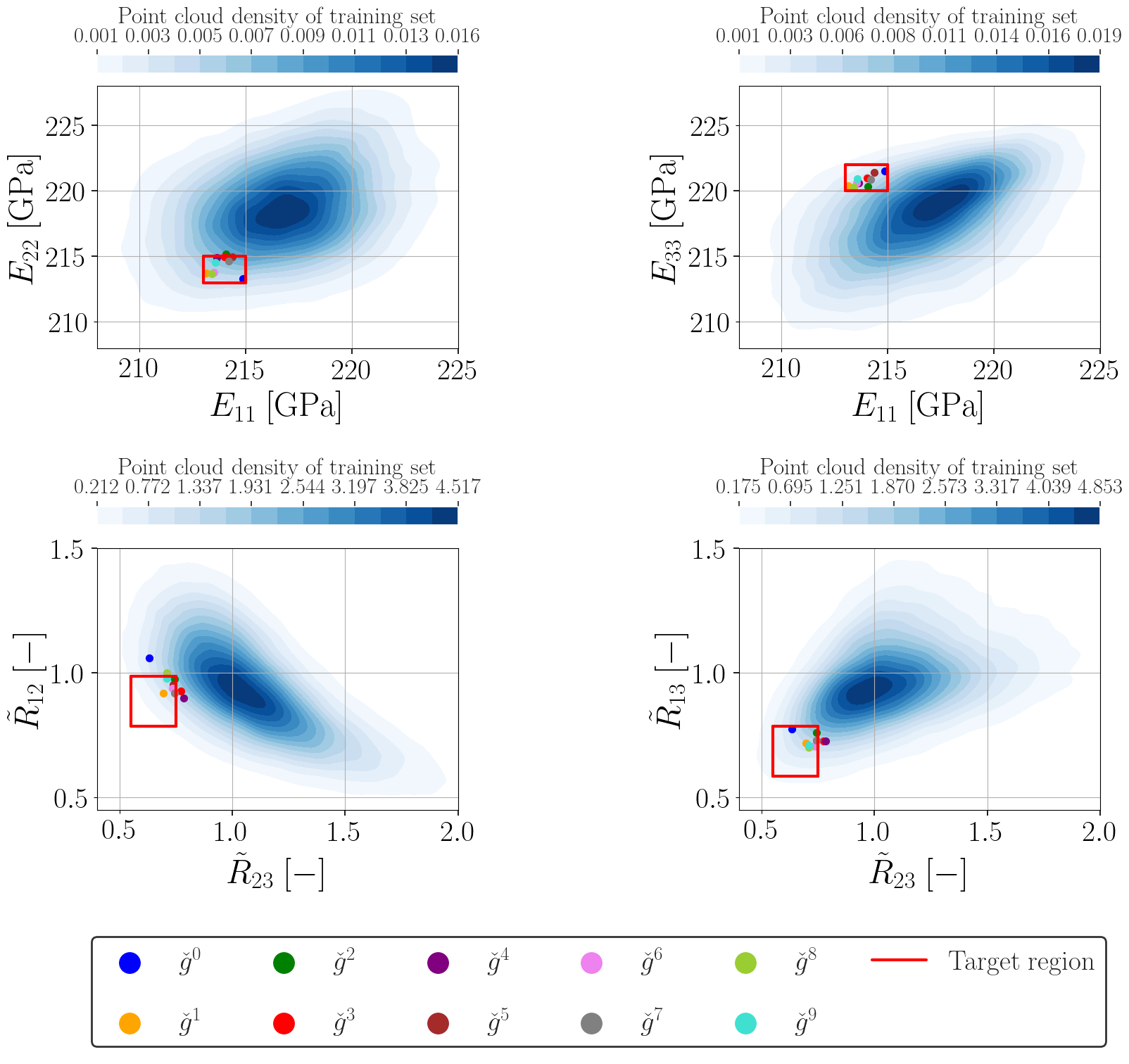}
\caption{Projections of the properties space showing the target region and the calculated properties of the ten chosen goal textures $\check{g}^0$ to $\check{g}^9$}
\centering
\label{fig:res_matdesign_goaltex}
\end{figure}

\begin{figure}
\centering
\includegraphics[width=0.6\linewidth]{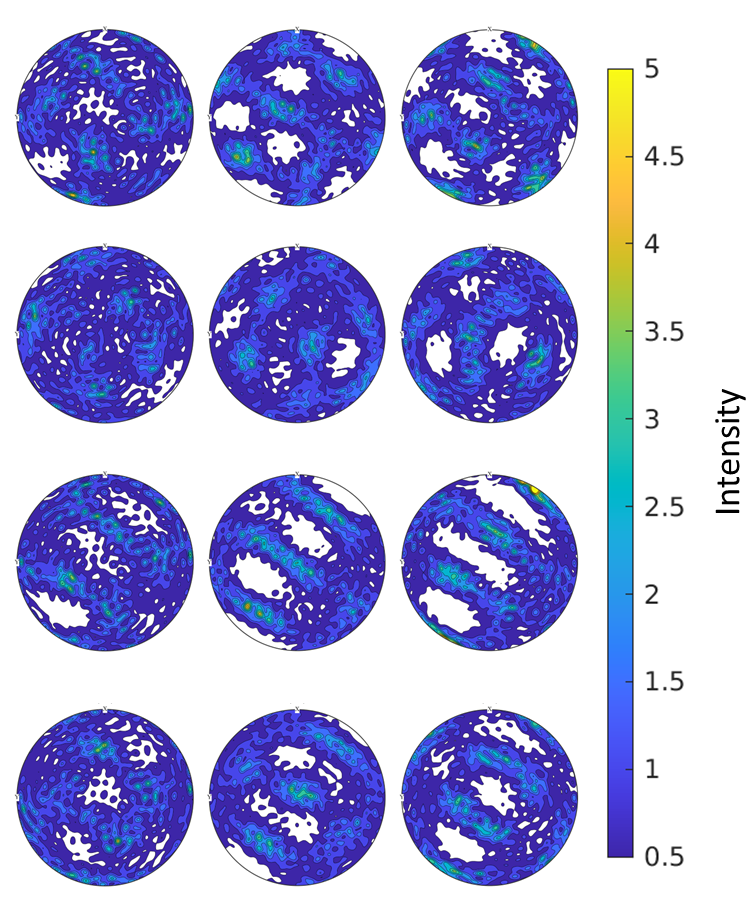}
\caption{Four exemplary crystallographic textures of the ten chosen goal textures depicted as pole figure plots: (001), (110), (111)}
\label{fig:res_matdesign_texexmpl}
\end{figure}

For the subsequent process design step, we chose ten goal textures from the set of identified crystallographic textures that differ strongly from each other \cite{Bauckhage2021-Max-Sum-Diversification}. The distribution of the chosen goal textures in properties space is shown in Figure \ref{fig:res_matdesign_goaltex}. The properties of the goal textures are mainly located inside the target region, with some exceptions that we tolerate for now and explain later in Section \ref{sec:discussion}. 
To show the differences between the goal textures, in Figure \ref{fig:res_matdesign_texexmpl}, four crystallographic textures are depicted exemplary as pole figure plots. While the pole figure intensities are similar for all crystallographic textures, the shape of the represented orientation distribution differs strongly.

The artificial neural networks and the reinforcement learning approach were trained a workstation equipped with 20 2.2 GHz CPU cores and a GeForce RTX 3090 GPU with 24GB RAM by using the Linux Ubuntu operating system.

\subsection{Process design using MEG-SGGPO} \label{sec:res_optproc}
Experiments are conducted with the following hyperparameters: Deep Q-learning as described in Section \ref{ss:meg-sggpo} as basic algorithm where the target-network is updated every $n_{\theta} = 50$ time-steps. Q-networks with hidden layer sizes of [128, 256, 256, 128], layer normalization and ReLU activation functions. The learning process starts after 100 control-steps. The networks are trained after each control-step with a mini-batch of size 32. The Adam optimizer \mbox{\cite{Kingma2015Adam}} is used for neural network training, with a learning rate of $5 \times 10^{-4}$. An $\epsilon$-greedy policy, with an initial exploration rate $\epsilon_0 = 0.5$ and the final exploration-rate $\epsilon_f = 0.0$, with $n_{\epsilon} = 390$. 

\begin{figure}
\centering
\includegraphics[width=0.8\linewidth]{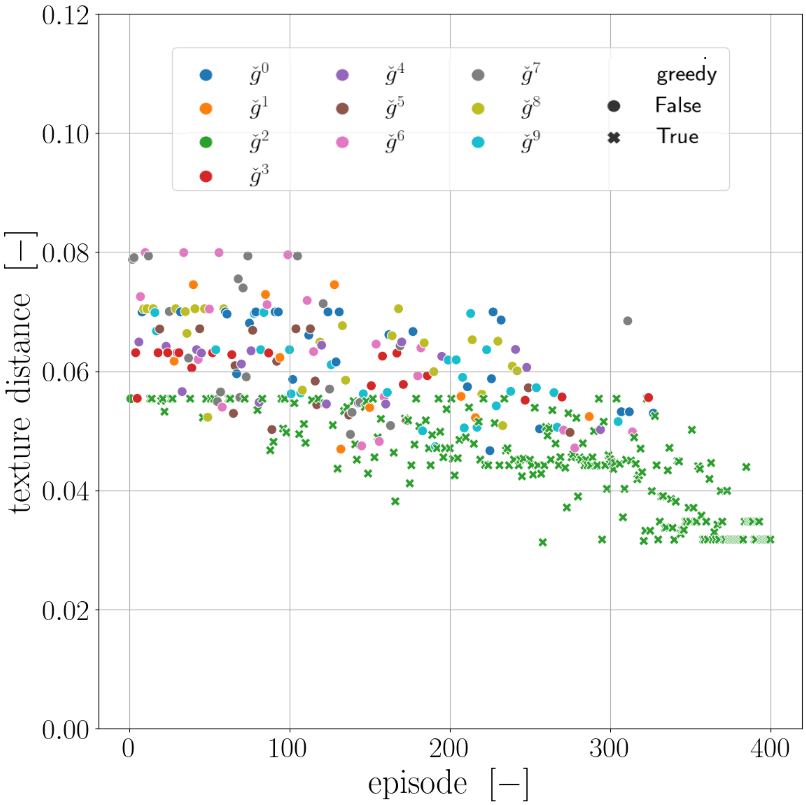}
\caption{Sinkhorn texture distance $\mathscr{D}_{\mathrm{sh}}$ between the produced and the goal texture over the episodes of the MEG-SGGPO approach. The color indicates the targeted crystallographic texture.}
\label{fig:results_RL_evolution}
\end{figure}

Once a diverse set of crystallographic textures has been identified, MEG-SGGPO is used to guide the metal forming process to the best reachable crystallographic texture. 
For solving the identification problem, the MEG-SGGPO approach is allowed to conduct $400$ process runs, so-called episodes. The evolution of the distance between the produced and the chosen goal texture is depicted  over the episodes in Fig. \ref{fig:results_RL_evolution}. Initially, the reinforcement learning agent attempts to identify processing paths that target randomly selected goal textures. As the learning process progresses, the agent focuses increasingly on producing goal texture $\check{g}^{2}$. 

\begin{figure}
\centering
\includegraphics[width=0.7\linewidth]{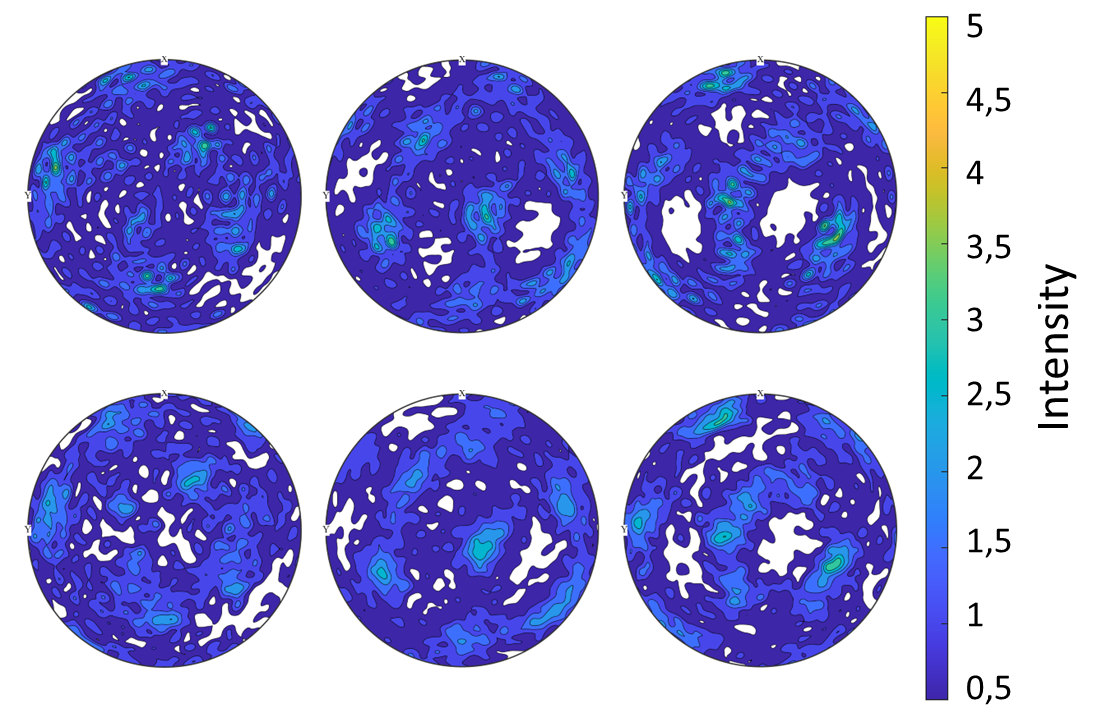}
\caption{Pole figure plots (001), (110), (111) of the targeted goal texture $\check{g}^{2}$ (top) and the produced crystallographic texture $\check{g}^{2'}$ (bottom)}
\label{fig:results_RL_reachedTex}
\end{figure}

For comparison, Figure \ref{fig:results_RL_reachedTex} presents pole figure plots of the goal texture $\check{g}^{2}$ and the produced crystallographic texture $\check{g}^{2'}$ after 400 episodes. Visually, the produced crystallographic texture is highly similar to the goal texture with intensity peaks at the same positions and of a similar magnitude. This also holds for areas not covered by orientations at all. To quantify the effectiveness of the approach, we compute the properties resulting from the produced crystallographic texture and compare them with the desired ones. The results, listed in Table \ref{tab:props_goal_reached_tex}, show that five of the properties ($E_{11}$, $E_{22}$, $E_{33}$, $\widetilde{R}_{12}$ and $\widetilde{R}_{13}$) are very close to the target region, while $\widetilde{R}_{23}$ lies inside. 

\begin{table}
\centering
\caption{Properties of the produced crystallographic texture and the distance to the target region}
\label{tab:props_goal_reached_tex}
\begin{tabular}{|c|c|c|c|}
\hline
 Target property\Tstrut\Bstrut & Value & Distance to target region & Unit \\
 \hline
$E_{11}$\Tstrut\Bstrut & 217.5 & 2.5 & GPa \\
$E_{22}$\Tstrut\Bstrut & 216.6 & 1.6 & GPa \\
$E_{33}$\Tstrut\Bstrut & 222.2 & 0.3 & GPa \\
\hline
$\widetilde{R}_{23}$\Tstrut\Bstrut & 0.744 & 0 & - \\
$\widetilde{R}_{12}$\Tstrut\Bstrut & 0.831 & 0.046 & - \\
$\widetilde{R}_{13}$\Tstrut\Bstrut & 1.058 & 0.073 & - \\
\hline
\end{tabular}
\end{table}

In summary, by sequentially applying the SMTLO and the MEG-SGGPO approach, we were able to successfully produce a crystallographic texture with desired properties in a simulated metal forming process. The resulting properties fall within the acceptable range from an engineering point of view, albeit at the border of the target region defined in properties space.

\section{Discussion}
\label{sec:discussion}

The results presented demonstrate the effectiveness of the SMTLO approach in identifying sets of near-optimal crystallographic textures that exhibit desired properties in a given target region. The set of crystallographic textures identified by the approach is shown to be more diverse than the benchmark set obtained from the training data set. However, the properties of the identified crystallographic textures are not completely inside the target region, as can be seen in Fig. \ref{fig:res_matdesign_goaltex}.   

There are two reasons for this: First, the formulation of the optimization objective in the SMTLO approach allows for a trade-off between (i) finding crystallographic textures with properties inside the target region, (ii) identifying possibly diverse crystallographic textures and, (iii) guaranteeing that crystallographic texture can be produced by the process under consideration. 
Second, due to prediction errors of the underlying machine learning model, the SMTLO approach identifies crystallographic textures whose predicted properties lie inside the target region, while some of the true properties (calculated by the numerical simulation on the basis of the reconstructed crystallographic textures) slightly lie outside. Both of these issues can be addressed by modifying the objective function of the SMTLO optimizer, defined by Equation \ref{eq:matdesign_obj}. For instance, targeting the center of the properties region, instead of its bounds can mitigate both issues. Additionally, the second issue can be addressed by enhancing the machine learning model, for example, by increasing the amount of training data. 

Taking a diverse subset of the identified crystallographic textures as goal textures, the MEG-SGGPO approach guides the metal forming process closely to one of the chosen goal textures. Although the reinforcement learning agent was not able to reproduce one of the goal textures exactly (which is challenging due to the many possible processing paths), the measured distance between the targeted crystallographic texture $\check{g}^{2}$ and the produced one $\check{g}^{2'}$  is sufficiently low ($\mathscr{D}_{\mathrm{sh}}=0.031$) when relating it to the distance to the nearest neighbor in the training data set ($\mathscr{D}_{\mathrm{sh}}=0.033$) and to the furthest data point ($\mathscr{D}_{\mathrm{sh}}=0.141$). This can be seen in Figure \mbox{\ref{fig:results_RL_evolution}}, which depicts the reduction in Sinkhorn distance from the initial texture $g_0$ to the final produced texture $\check{g}^{2'}$ guided by the reinforcement learning agent. 

We want to remark here, that it is generally difficult to evaluate distances between crystallographic textures, even for experts. Therefore, the quality of the produced crystallographic texture is determined by two considerations: (i) the distance of the textures in the microstructure space (as illustrated above), which is statistically valid, and (ii) the distance in properties, which can be assessed based on the results shown in Table \mbox{\ref{tab:props_goal_reached_tex}}. It can be seen that for the produced texture, one property lies within the target region, while the remaining five properties are in close proximity. It is remarkable that it took the reinforcement learning agent only 400 episodes to achieve this result compared to the baseline set that is grounded on the generation of 76980 random samples. The reinforcement learning algorithm can therefore be seen as being data efficient.

Nevertheless, in this study, it seems that a lower bound is existing that is difficult to overcome by the MEG-SGGPO approach. In general, this can have two reasons: 
First, the forming process is unable to produce crystallographic textures that are closer to the goal texture identified by the SMTLO approach. Yet, the SMTLO approach already addresses this issue by enforcing microstructures to remain within the region delineated by the known microstructures from the training data set via the term $\mathscr{V}_{\mathrm{valid}} \mathscr{C}_{\mathrm{valid}}$ in Equation (\ref{eq:matdesign_obj}). We expected that a stronger enforcement leads to crystallographic textures that are better reachable by the process. This, however, can lead to a lower accuracy in terms of obtaining the desired material properties, due to the change in the other weights in the objective function of the SMTLO optimizer. Second, the MEG-SGGPO approach identified a local optimum and got stuck. This can be mitigated by longer MEG-SGGPO runs with optimized hyperparameters. Longer runs, however, have not shown to yield significant improvements in the presented study. 
As future work, it is desirable to incorporate the knowledge contained in the generated training samples for SMTLO into MEG-SGGPO a priori to enhance its performance. 

\section{Conclusion}
\label{sec:conclusion}

In summary, the machine learning-based approach presented in this study enables accelerated materials development by taking into account details of the manufacturing process. This was particularly challenging as for the considered metal forming process a total of $25^7$ different processing paths would be possible; we found that
\begin{itemize}
\item $76980$ samples were sufficient for the SMTLO approach to successfully solve the materials design problem. It allowed us to identify a diverse set of near-optimal crystallographic textures with desired properties. The crystallographic textures are considered near-optimal in the sense that all of them bear different but satisfactory macroscopic properties and, at the time point of the identification, it is not know which one is best reachable by the process.
\item after solving the materials design problem, a total of $400$ episodes was sufficient for the MEG-SGGPO approach to successfully guide the forming process to the best reachable crystallographic texture and solve the process design problem.
\end{itemize}

This shows that the applied approach is highly data efficient and capable of effectively optimizing process-structure-property relations end-to-end in manufacturing processes. Moreover, we emphasize that the approach leverages the non-uniqueness of the materials and process design problems, i.e. a diverse set of microstructures is identified for given desired properties and subsequently, the best reachable microstructure is produced. This is a significant advantage when transferring the approach to real manufacturing systems, where often constraints exist, which may exclude microstructures from being producible. 

Nonetheless, the application still needs to be validated in real manufacturing systems. In our specific case, the amount of required training data is relatively high as compared to what is available in typical production workflows. Thus, for application in real-world manufacturing, a future scenario could involve pre-training the machine learning models with a numerical simulation that serves as a digital twin of the process and the material.

\section*{Data availability statement}
Data and code has been published in the following repositories:
\begin{itemize}
\item Training data for the SMTLO approach is made available via Fraunhofer Fordatis repository at \url{https://fordatis.fraunhofer.de/handle/fordatis/319} \cite{processdata2023}.
\item The process simulation is made available via Fraunhofer Fordatis repository at \url{https://fordatis.fraunhofer.de/handle/fordatis/201.2}\cite{processsim}.
\item The source code for the Siamese multi-task learning framework is made available via gitlab at \url{https://gitlab.com/tarekiraki/the_smtl_framework}.
\item The reinforcement learning environment is made available via github at \url{https://github.com/Intelligent-Systems-Research-Group/RL4MicrostructureEvolution}.
\end{itemize}

\section*{Acknowledgments}
The authors would like to thank the German Research Foundation (DFG) for funding this work, which was carried out within the research project number 415804944: ’Taylored Material Properties via Microstructure Optimization: Machine Learning for Modelling and Inversion of Structure-Property-Relationships and the Application to Sheet Metals’.

\section*{Author contribution}
LM set up the numerical simulation, generated the training data for solving the materials design problem and wrote the manuscript together with all other authors. 
TI developed the SMTLO approach and applied both, the SMTLO and the MEG-SGGPO approach to the problem analyzed in this study. 
JD developed the MEG-SGGPO approach. The approach was extended using the Sinkhorn distance measure by TI.
SS supported in terms of machine learning and materials science.
NL supported in terms of machine learning.
DH supported in terms of materials science. All authors contributed equally to the discussion of the results and to reviewing the manuscript. 


\section*{Declaration of generative AI and AI-assisted technologies in the writing process}

During the preparation of this work the authors used ChatGPT in order to improve language and readability. After using this tool, the authors reviewed and edited the content as needed and take full responsibility for the content of the publication.

\bibliographystyle{ieeetr}
\bibliography{literature}

\end{document}